\documentclass[aps,pre,preprint,10pt,superscriptaddress,showpacs]{revtex4-1}
\usepackage{amsfonts} 
\usepackage{amsmath}
\usepackage{amssymb}
\usepackage{graphicx}%\[  \]
\usepackage{subfigure}
\usepackage{color}
\newcommand*\xbar[1]{%
  \hbox{%
    \vbox{%
      \hrule height 0.5pt % The actual bar
      \kern0.5ex%         % Distance between bar and symbol
      \hbox{%
        \kern-0.1em%      % Shortening on the left side
        \ensuremath{#1}%
        \kern-0.1em%      % Shortening on the right side
      }%
    }%
  }%
} 
\begin{document}
\title{Drift ion-acoustic waves in a nonuniform rotating magnetoplasma with two-temperature superthermal electrons}
\author{A. Abdikian}
\email{abdykian@gmail.com}
\affiliation{Department of Physics, Malayer University, Malayer 65719-95863, Iran}
 \author{M. Eghbali}
 \email{eghbali.mohammad57@gmail.com}
 \affiliation{Department of physics, Faculty of science, Behbahan Khatam Alanbia University of Technology, Behbahan, 63616-47189, Iran}
\author{A. P. Misra}
\email{apmisra@visva-bharati.ac.in; apmisra@gmail.com}
\affiliation{Department of Mathematics, Siksha Bhavana, Visva-Bharati University, Santiniketan-731 235, West Bengal, India}
%%%%%%%%%%%%%%%%%%%%%%%%%% 
\begin{abstract}
The theory of low-frequency (in comparison with the ion cyclotron frequency), long wavelength,  electrostatic drift ion-acoustic waves (IAWs) is studied in a nonuniform rotating magnetoplasma with two temperature superthermal electrons. In the linear limit, the coupling of IAWs and drift waves by the density inhomogeneity is shown to produce  a new wave mode which typically depends on the density gradient, the rotational frequency and the spectral indexes of superthermal electrons. In the nonlinear regime,  an evolution equation for the drift IAWs is derived by the dispersion approach, and using the Jacobi elliptic function expansion technique its exact solitary and periodic wave solutions (namely, cnoidal and dnoidal) are also obtained. The properties of these solutions are numerically examined and it is found that they are significantly modified by  the effects of the background density gradient, the superthermality of electrons and the Coriolis force associated with the rotational motion of ions.    
      
\end{abstract}
\maketitle
%\begin{keywords}
%Rotating plasma, Drift waves, Superthermal electrons, the modified Zakharov-Kuznetsov equation.
%\end{keywords}

\section{Introduction} \label{sec-intro}

The  low-frequency ion-acoustic waves (IAWs) have been one of the most fundamental electrostatic oscillations  in multi-component plasmas and a topic of foremost research over the past many years \cite{Krall, Rao375, Barkan239}.  Such waves are ubiquitous not only in laboratories but also in various planetary and space environments including those in   auroral zone \cite{Olsson298}, in the Earth's magnetosphere \cite{Christon1}, in the interstellar medium \cite{Leubner547},  the solar wind \cite{Shrauner123}  and cometary tails \cite{scarf1986}.  One of the nonlinear evolution equations  which governs the dynamics of multi-dimensional electrostatic waves in a magnetoplasma is the Zakarov-Kuznetsov (ZK) equation \cite{Zakharov594}. In recent times, the latter  has been studied by many authors in various plasma environments \cite{Mace2649, Mushtaq072306, Abdikian122303, Mahmood035505}.  Furthermore, many laboratory experiments \cite{liu1994} and  spacecraft observations (e.g., in the vicinity of the Moon,  the upper ionosphere of Mars, Earth’s bow shock, etc.) \cite{maksimovic1997}  validate the existence 
of highly energetic   superthermal particles which exhibit some deviation from the thermodynamic equilibrium and so their   distributions cannot be described by the  Maxwellian-Boltzmann distribution. 
A lot of attention has been paid to investigate the influence of these energetic particles on the propagation of electrostatic or electromagnetic waves in recent times due to their potential applications in laboratory \cite{Hellberg433, Goldman145002} and space \cite{Vasyliunas2839, Hasegawa2608, Pierrard7923, PierrardA02118, Abdikian095602} plasma environments.  Although, several models have been proposed, one known distribution function  describing the energetic superthermal particles  is the kappa distribution function \cite{Livadiotis}. 
The form of such distribution function in space plasmas was first proposed by Vasyliunas \cite{Vasyliunas2839}, which can be written as  \cite{BalukuA04227, Pierrard153}
\begin{equation}
f_\kappa(v)=\frac{n_{0}}{\left(\pi\kappa\theta^2\right)^{3/2}}\frac{\Gamma(\kappa+1)}{\Gamma(\kappa-1/2)}\left(1+\frac{v^2}{\kappa\theta^2}\right)^{-(\kappa+1)},
\end{equation}
where $\theta=\left\{\left(\kappa-3/2\right)/\kappa\right\}^{1/2}v_\text{th}$ represents the effective particle velocity related to the  particle's thermal velocity $v_\text{th}=\left(2k_BT_e/m_e\right)^{1/2}$ with  $T_e~(m_e)$ denoting the   temperature (mass)  of electrons, $k_B$   the  Boltzmann constant;  $\Gamma$ is the Gamma Function, $n_{0}$ is the equilibrium number density  and $\kappa$ is  the spectral index. Here, the parameter $\kappa$ is a measure of the slope of the energy spectrum of  superthermal particles in the distribution function. The smaller the values of $\kappa$,   the larger is the number of particles that deviate from the Maxwell distribution $(\kappa\rightarrow\infty)$   and that are placed at the tail of the distribution.   
\par
   Krall and Rusenbluth \cite{Krall254}   showed that  in addition to purely transverse unstable modes, there also exist stable longitudinal modes   in a nonuniform magnetoplasma.  The latter with a nonuniform background magnetic field can  support  a great variety of low-frequency (in comparison with the cyclotron frequency) electrostatic    drift  modes \cite{Rudakov415,D'Angelo422}, e.g.,  the electrostatic coupled drift Alfv{\'e}n modes \cite{misra2009}, electrostatic drift wave envelopes \cite{shukla2012}, which play significant roles in particle confinement in Tokomaks and Q machines \cite{Horton735}; cross-field plasma particle transports \cite{horton1999},  as well as, the formation of nonlinear coherent structures  \cite{shukla2012} in laboratory and space plasmas.    
\par 
Various interesting phenomena can also occur when the Coriolis force due to the Earth’s rotation  is taken into account in the fluid motion of plasma particles. It has been shown that the Coriolis force not only couples the  high- and low-frequency  acoustic-like waves  but also modifies the resonance and cut-off frequencies of various   atmospheric waves \cite{chatterjee2021}. Furthermore, depending on the parameter regimes both the Lorentz force and the Coriolis force (or any one of  them) can influence the evolution of nonlinear waves \cite{kaladze2008}. For example,  in the Earth's ionospheric D-layer, the Lorentz force contribution is negligible compared to the Coriolis force but the same can dominate over the Coriolis force in the Earth's F-layer. However, both the forces can have influence in the E-layer in which the effects of the magnetic field corresponds to a replacement of the rotational frequency $2\Omega_0$ by $2\Omega_0+\omega_{ci}$, where $\omega_{ci}$ is the ion cyclotron frequency. 
 Thus, the properties of drift waves along with their coupling   with ion-acoustic waves remain one of the most promising areas of research   \cite{Mushtaq042305, Kourakis018, Haque092102, Farooq122} and   a topic of current interest.
\par
The objective of this work is to explore the  linear and nonlinear theories of low-frequency electrostatic drift  IAWs and the formation of solitary and periodic wave structures in a nonuniform  rotating magnetoplasma consisting of two groups of kappa distributed superthermal electrons  with  different thermal energies and positive ions.     It is pertinent to mention that before the thermal equilibrium is reached, there must exist a time  scale in which  the division of  electrons into two groups with different temperatures and number densities is possible. Such a time scale ($\tau$) should be longer than the the typical time scale for the drift ion-acoustic waves to exist, which is typically $\sim (1/\omega_{ci})\sim 10^{-5}$ s for the parameters used in Sec. \ref{5}. Here, $\omega_{ci}\equiv eB_0/cm_i$ is the ion cyclotron frequency.  For a Maxwellian plasma, the typical time for two populations (having the same mass $m_e$) to equilibrate, i.e., the time of equipartition is given by \cite{spitzer1962}
\begin{equation}
\tau_\mathrm{eq}=\frac{3\sqrt{m_e}\left(k_BT_l+k_BT_h\right)^{3/2}}{8\sqrt{2\pi} n e^4 \ln\Lambda},
\end{equation}  
where $\ln \Lambda$ is the collisionality parameter (Coulomb logarithm).  It has been estimated that $\tau_\mathrm{eq}\sim 6$ s if the mean square relative velocity $\sim k_B(T_l+T_h)/m_e$ of the two species does not change appreciably, and so $\tau_\mathrm{eq}\gg\omega_{ci}^{-1}$. Thus, for a superthermal plasma with a tail of hot species having  $n_{h0}>n_{l0}$ and $T_h>T_l$, it is expected that the typical time of thermalization of electrons is much longer than the time scale of ion oscillations.  
On the other hand, there might be a situation where the unperturbed plasma corresponds to an electron fluid streaming through the neutralizing background with a nonzero velocity. In this case, the instability of ion-acoustic waves  may occur that can lead to the thermalization of low-temperature species. However, this case is not considered in the present study on the assumption that the ion-acoustic phase velocity is larger than the streaming velocity of plasma particles.   
\par 
The paper is organized as follows.   In Section \ref{2}, the set of basic   equations is presented for the dynamics of magnetized cold ions and two groups of superthermal low- and high-temperature electrons.   Section \ref{3} presents the derivation of the linear dispersion relation and the description of drift and ion-acoustic modes.    The modified ZK-like equation governing   the evolution of drift IAWs is derived in Sec. \ref{4} following the dispersion relation approach \cite{Gell402}.   Section \ref{5} is left for the numerical study of the characteristics of drift wave modes and exact solutions of  the ZK-like equation including the traveling and periodic solutions. The latter  are obtained by using the Jacobi elliptic function expansion method. Finally, Sec. \ref{6} concludes the results. 
%%%%%%%%%%%
\section {Basic equations}\label{2}
We consider the propagation of low-frequency (in comaprison with the ion cyclotron frequency) ion-acoustic waves that are coupled to the drift waves in a collisionless, nonuniform, magnetized   plasma with two groups of  superthermal electrons (namely, low- and high-temperature energetic electrons). The uniform external magnetic field is embedded along the $z$-axis, i.e.,  $ \mathbf{B}_0=B_0\hat{z}$. We assume that the background number density of  plasma particles has the gradients   along the $x$-axis such that $ \nabla n_{j0}(x)\sim dn_{j0}(x)/dx<0$. Furthermore, the ions are subjected to the  Coriolis force  which appears due to the Earth's rotation with    angular velocity $ \mathbf{\Omega}=\Omega_0\hat{\textbf{z}}$ along the $z$-axis. The basic  equations for ion fluids are \cite{Adnan092119, Farooq122110, Ahmad905810505} 
\begin{eqnarray}
	&&\frac{\partial n_i}{\partial t}+{\mathrm{\nabla}}\cdot\left(n_i {\mathbf{u}_i}\right)=0,\label{e1}\\
	&&\frac{\partial {\mathbf{u}_i}}{\partial t}+\left({\mathbf{u}_i}\cdot\mathrm{\nabla}\right) {\mathbf{u}_i}=\frac{e}{m_i}\left( -\nabla\phi+\frac{1}{c} {\mathbf{u}_i}\times {\mathbf{B}}\right)+2\Omega_0({{\mathbf{u}}}_i\times\hat{\mathbf{z}}),\label{e2}\\
	&&\mathrm{\nabla}^2 \mathrm{\phi}=4\pi e (n_h+n_l-n_i),\label{e3}
	\end{eqnarray}
	and those for superthermal electrons are \cite{BalukuA04227, Pierrard153}
	\begin{eqnarray}
	&&n_h=n_{h0}\left[1-\frac{e\sigma\phi}{T_h\left(\kappa_h-3/2\right)}\right]^{-\kappa_h+1/2},\label{e4}\\
	&&n_l=n_{l0}\left[1-\frac{e\phi}{T_l\left(\kappa_l-3/2\right)}\right]^{-\kappa_l+1/2},\label{e5}
\end{eqnarray}
where $n_i$, $m_i$ and ${\mathbf{u}}_i$ are, respectively,  the     number density, mass and fluid velocity of ions;  $ \phi$ is the  electrostatic potential,  $e$ is the elementary charge, $ c$ is the speed of light in vacuum and   $\sigma=T_l/T_h$ is the   ratio between the temperatures low- (with suffix $l$) and high- (with suffix $h$) temperature   superthermal electrons.  We assume that the rotational frequency is small, i.e., $\Omega_0<\partial/\partial t$  for which  one may ignore the  terms involving second and higher order of $\Omega_0$ and hence the effect of the centrifugal force $ {\mathbf{\Omega}}_0\times({{\mathbf{\Omega}}}_0\times{\mathbf{r}})$ in the ion motion \cite{Abdikian095605}. In astrophysical plasmas, Chandrasekhar suggested that not only the magnetic field influences the propagation dynamics of electrostatic waves but also the Coriolis force due to the Earth's rotation that plays a key role in the cosmic processes \cite{Farooq122110, Chandrasekhar667}. So, in the rotating frame of reference, the Coriolis force  should be added to the fluid equation of motion \eqref{e2}.  The   derivation of the Coriolis force is presented in Appendix A.  
In what follows, the charge neutrality condition at equilibrium reads $n_{l0}+n_{h0}=n_{i0}$, i.e., $\mu\equiv n_{h0}/n_{l0}=n_{i0}/n_{l0}-1$  where $n_{j0}$ stands for the equilibrium number density of $j$-th species particles (with the suffix $j=l$  for low-temperature electrons, $j=h$  for high temperature electrons  and $j=i$  for cold ions). 
\par 
 In the small amplitude limit of electrostatic perturbations, i.e.,  $e\phi/T_l\ll1$, one can expand   Eqs. \eqref{e4} and \eqref{e5} in powers of $\phi$.  Substituting these expansions into   Eq. \eqref{e3} we obtain  
\begin{equation}
	\left(\gamma-\lambda_{Dl}^2\mathrm{\nabla}^2\right)\phi=4\pi e\ \lambda_{Dl}^2n_i. \label{e6}
\end{equation}
where $\lambda_{Dl}=\sqrt{ {T_l}/{4\pi n_{l0}e^2}}$ is the Debye length of low-temperature electrons and $\gamma=\beta_l+ \mu\sigma\beta_h$ with  $\beta_j=(\kappa_j-1/2)/(\kappa_j-3/2)$, $j=l,h$, represents the parameter involving the spectral indexes of superthermal species. 
\section{Dispersion relation}\label{3}
To obtain the linear dispersion relation for the low-frequency ($\left|\partial/\partial t\right|\ll\omega_{ci}$)  drift IAWs, we linearize the basic equations \eqref{e1} to \eqref{e5} about the equilibrium state $(n_{j0},0,0)$ of ($n_j,~u_i$, $\phi$), and assume  the perturbations to vary as plane waves $\sim\exp(i\mathbf{k}\cdot \mathbf{r}-i\omega t)$  with the wave vector $\mathbf{k}=(k_y,k_z)$ and wave frequency $\omega$.   Here, we note that although   in rotating fields  the cylindrical geometry may be the right choice, the Cartesian geometry can still be considered for the plane wave  perturbations. The reason is that for a plane wave perturbation of the form $f(\mathbf{r},t)=\tilde{f}\exp(i\mathbf{k}\cdot\mathbf{r}-i\omega t)$ with constant amplitude $\tilde{f}$, the time derivative of $f$ in the rotating frame contributes a factor $i(\mathbf{k}\cdot{d\mathbf{r}}/{dt}-\omega)\sim i(\mathbf{k}\cdot \mathbf{\Omega}\times \mathbf{r}-\omega)\sim i(\Omega_0 k_y y-\omega)$. For  long-wavelength perturbations with $k^2_y\rho_i^2\ll1$ and $\Omega<\omega$, the term $\propto\Omega_0$ can be disregarded compared to $\omega$ to leave only the contribution $-i\omega$. So, our consideration of using the Cartesian geometry is still valid. From Eq. \eqref{e2}, the perturbed velocity components are  
\begin{equation}\label{e7}
	u_{ix}=-\left(\frac{e}{m_i\Omega_c}\right)\frac{\partial\phi}{\partial y},\ \ u_{iy}=i\left(\frac{e\omega}{m_i\Omega_c^2}\right)\frac{\partial\phi}{\partial y},\ \ u_{iz}=-\left(\frac{i\ e}{m_i\omega}\right)\frac{\partial\phi}{\partial z},
\end{equation}
where $\Omega_c=\omega_{ci}+2\Omega_0$ is the effective frequency modified the Coriolis force. 
\par 
Next, the linearization of the ion continuity equation \eqref{e1}  gives
\begin{equation}\label{e8}
	\frac{\partial n_i}{\partial t}+n_{i0}u_{ix}K_{ni}+n_{i0}\frac{\partial u_{iy}}{\partial y}+n_{i0}\frac{\partial u_{iz}}{\partial z}=0,  
\end{equation}
where $K_{ni}=\left|(1/n_{i0}) (\partial n_{i0}/\partial x)\right|$ denotes the inverse of the density inhomogeneity scale length. Eliminating $u_{iy}$ and $u_{iz}$ from Eqs. \eqref{e7} and \eqref{e8}, we obtain
\begin{eqnarray}
	\left(\gamma-\lambda_{Dl}^2\mathrm{\nabla}^2\right)\phi=\frac{c_s^2}{\omega^2\Omega_c^2}\frac{n_{i0}}{n_{c0}}\left[\omega^2\frac{\partial^2\phi}{\partial y^2}-i K_{ni}C_s\omega\frac{\partial\phi}{\partial y}-\Omega_c^2\frac{\partial^2\phi}{\partial z^2}\right],\label{e9}
\end{eqnarray} 
where $c_s=\sqrt{T_l/m_i}$ is the ion-acoustic speed for a simple unmagnetized Maxwellian electron-ion plasma.  Next, Fourier analyzing Eq. \eqref{e9}  we obtain the following linear dispersion relation  for  coupled drift and ion-acoustic waves in   a nonuniform superthermal magnetoplasma. 
\begin{eqnarray}
	\left[\gamma+\lambda_{Dl}^2k^2+(1+\mu)\rho_i^2k_y^2\right]\omega^2-(1+\mu)\omega_\ast\omega-(1+\mu)c_s^2k_z^2=0,\label{e10}
\end{eqnarray}
where $\rho_i=c_s/\Omega_c$ is the effective gyro-radius, $\omega_\ast=-u_d^\ast k_y$ with $u_d^\ast= {T_lK_{ni}}/{m_i\Omega_c}$ denoting the modified diamagnetic drift velocity and $k^2=k_y^2+k_z^2$. From Eq. \eqref{e10} it is evident that the coupled drift IAWs is significantly modified by the effects of the density inhomogeneity, the superthermality of electrons as well as the Coriolis force. In fact, the inclusion of two groups of superthermal electrons modifies the Debye screening length, and hence the dispersion due to separation of charges. Here, we note that in absence of the density inhomogeneity, the drift mode and the IAW are decoupled.  We will numerically investigate the dispersion properties of the drift IAWs shortly in Sec. \ref{5}.  
\par 
Some particular cases may be of interest:
\begin{itemize}
\item  In the long-wavelength limit ($\lambda_{Dl}^2k^2,\rho_i^2k_y^2\ll1$) and for a two-component electron-ion plasma (for which $\mu=0$), the dispersion equation \eqref{e10} reduces to the known form \cite{Gell402}
\begin{equation}
\omega^2-\omega_\ast\omega-c_s^2k_z^2=0.
\end{equation}
\item When $c_s^2k_z^2\gg\omega_\ast^2$, the two branches  of solutions of Eq. \eqref{e10} reduce to the ordinary IAW, given by,
 \begin{eqnarray}
	 \omega^2=\frac{c_s^2k_z^2\left(1+\mu\right)}{\gamma+k^2\lambda_{Dl}^2+(1+\mu)\rho_i^2k_y^2}. \label{eq-iaw} 
\end{eqnarray}
Here, in absence of the Coriolis force and  the density inhomogeneity, the long-wavelength   ion-acoustic mode can be recovered from Eq. \eqref{eq-iaw} as in Refs. \cite{Panwar122105, Bains1}, i.e., 
 \begin{eqnarray}
	\omega^2=\frac{c_s^2k_z^2\left(1+\mu\right)}{\beta_l+\mu\sigma\beta_h}.\label{e11}
\end{eqnarray}
%From Eq. \eqref{e11}, it follows that while the IAW frequency increases with increasing values of the density ratio $\mu$, it decreases with increasing values of either the temperature ratio $\sigma$ or the spectral indices $\beta_l$ and $\beta_h$. 
\item As one approaches from a domain of higher to lower values of $k_z$, the upper branch of the wave modes, given by Eq. \eqref{e10}, disappears from the IAW and approaches to $\omega_\ast$ as $k_z\rightarrow0$. In this case,   the high-frequency drift mode with $\omega^2\gg c_s^2k_z^2$  can be obtained  as  
\begin{eqnarray}\label{e12}
	\omega=\frac{\omega_\ast(1+\mu)}{\gamma+\lambda_{Dl}^2k^2+(1+\mu)\rho_i^2k_y^2}.
\end{eqnarray}
\end{itemize}
From the above analysis, it is expected that in the intermediate region between the  two limits (given in the last two items), the nonlinear evolution equation of the drift IAWs may  be similar to that governs the ordinary IAWs [e.g., Korteweg-de Vries (KdV) equation in one dimension and ZK equation in multi dimensions]. In Sec. \ref{4}, we consider the nonlinear evolution of drift IAWs in the intermediate regime. 
\section{Nonlinear evolution equation}\label{4}
We  derive an evolution equation for drift IAWs using the dispersion approach \cite{Gell402}. To this end, we first note that in the limits of  $\lambda_{De}^2\mathrm{\nabla}^2/\gamma$, $\omega^2/\Omega_c^2$, $\left|K_{ni}k_y\omega/(k_z^2\Omega_c)\right|\ll1$,   Eq. \eqref{e9} gives as a first approximation  
\begin{eqnarray}\label{e13}
	\omega^2\phi=-\frac{c_s^2}{\gamma}(1+\mu)\frac{\partial^2\phi}{\partial z^2}.
\end{eqnarray}
Inserting this result into Eq. \eqref{e9} we get
\begin{eqnarray}\label{e14}
	\left(\gamma-\lambda_{De}^2\mathrm{\nabla}^2\right)\phi=-\frac{c_s^2}{\omega^2}(1+\mu)\left[(1+\mu)(1+\mu)\frac{\rho_i^2}{\gamma}\frac{\partial^4\phi}{\partial y^2\partial z^2}+\frac{\partial^2\phi}{\partial z^2}+\frac{K_{ni}}{\Omega_c}\frac{c_s}{\sqrt\gamma}\sqrt{1+\mu}\frac{\partial^2\phi}{\partial y\partial z}\right],
\end{eqnarray}
from which  one obtains  
\begin{eqnarray}\label{e15}
	\omega\approx\frac{c_sk_z}{\sqrt\gamma}\sqrt{1+\mu}\left[1-\frac{1}{2\gamma}\left(\lambda_{De}^2+(1+\mu)\rho_i^2\right)k_y^2-\frac{1}{2\gamma}\lambda_{De}^2k_z^2\right]+\frac{1+\mu}{2\gamma}u_d^\ast k_y.
\end{eqnarray}
The corresponding equation of Eq. \eqref{e15} in terms of the variables $z$ and $t$,    when applied to the third relation of Eq. \eqref{e7}, gives the following relation.
\begin{eqnarray}\label{e16}
	\frac{e}{m_i}\frac{\partial\phi}{\partial z}=\frac{C_s\sqrt{1+\mu}}{\sqrt\gamma}\left[\frac{\partial u_{iz}}{\partial z}+\left(\frac{\lambda_{Dl}^2+(1+\mu)\rho_i^2}{2\gamma}\right)\frac{\partial^3u_{iz}}{\partial z{\partial y}^2}+\frac{1}{2\gamma}\lambda_{Dl}^2\frac{\partial^3u_{iz}}{{\partial z}^3}\right]+\frac{1+\mu}{2\gamma}u_d^\ast\frac{\partial u_{iz}}{\partial y}.
\end{eqnarray}
From Eq. \eqref{e16} and  the $z$-component of the momentum equation \eqref{e2}  we obtain the following ZK-like equation for the evolution of low-frequency drift IAWs in a nonuniform superthermal magnetoplasma. 
%\begin{eqnarray}\label{e17}
%	\frac{\partial u_{iz}}{\partial t}+u_{iz}\frac{\partial u_{iz}}{\partial z}+\frac{C_s\sqrt{1+\mu}}{\sqrt\gamma}\left[\frac{\partial u_{iz}}{\partial z}+\left(\frac{\lambda_{De}^2+(1+\mu)\rho_i^2}{2\gamma}\right)\frac{\partial^3u_{iz}}{\partial z{\partial y}^2}+\frac{\lambda_{De}^2}{2\gamma}\frac{\partial^3u_{iz}}{{\partial z}^3}\right]+\frac{1+\mu}{2\gamma}u_d^\ast\frac{\partial u_{iz}}{\partial y}=0.
%\end{eqnarray}
\begin{eqnarray}\label{e18}
	\frac{\partial u_{iz}}{\partial t}+u_{iz}\frac{\partial u_{iz}}{\partial z}+A\frac{\partial u_{iz}}{\partial z}+B\frac{\partial^3u_{iz}}{\partial z{\partial y}^2}+C\frac{\partial^3u_{iz}}{{\partial z}^3}+D\frac{\partial u_{iz}}{\partial y}=0.
\end{eqnarray}
Here, we have  introduced the dimensionless variables according to $u_{iz}\rightarrow u_{iz}/c_s$, $(y,z)\rightarrow (y,z)/\rho_i$ and $t\rightarrow\Omega_ct$. The coefficients appearing in Eq. \eqref{e18} are 
\begin{eqnarray}\label{e19}
	A=\frac{\sqrt{1+\mu}}{\sqrt\gamma},\ \ B=\frac{A}{2\gamma}\left(1+\mu+\frac{\lambda_{Dl}^2}{\rho_i^2}\right),\ \ C=\frac{A}{2\gamma}\frac{\lambda_{Dl}^2}{\rho_i^2},\ \ D=\frac{(1+\mu)u_d^\ast}{2\gamma\ c_s}.
\end{eqnarray}
In comparison of Eq. \eqref{e18}  with the usual ZK equation \cite{Zakharov594}, it is noted   that there are some additional terms $\propto A$ and $D$. While the former appears due to the drift wave approximation, the latter is associated with the density inhomogeneity. Considering the wave motion in a frame moving with the speed $c_s\sqrt{(1+\mu)/\gamma}$ and the uniform density (for which $D$ vanishes) one can recover the known form of the ZK equation. 
 From Eq. \eqref{e18} we also note that   both the nonlinear and dispersion terms are  modified by the effects of the density inhomogeneity, the superthermality of electrons as well as the influences of the Lorentz and Coriolis forces, causing to  significantly modify the profiles of the nonlinear coherent structures that can be formed. 
\par 
It is imperative to note that the ZK-like equation \eqref{e18} can support various interesting  solitary and periodic wave solutions.
  In a moving frame $\xi=\lambda y+z-u_0t$ with constant velocity $u_0$ and $\lambda<1$, using the Jacobi elliptic function expansion method \cite{Abdikian095602, Liu69, Fan6853, Abdikian997, Fan819, Zhang144, Prasad, Abdikianwaves} the solitary wave solution can be obtained as
\begin{eqnarray}\label{e20}
	u_{iz}=12\left(B\lambda^2+C\right) \mathrm{sech}^2  \left(\lambda y+z-u_0t\right),
\end{eqnarray}
where $u_0=A+D\lambda+4(C+B\lambda^2)$. Furthermore, we obtain  the cnoidal and dnoidal solutions of  Eq. \eqref{e18}  as
\begin{eqnarray}\label{e21}
	u_{iz}=\frac{12\left(B\lambda^2+C\right)}{2m^2-1}\mathrm{cn}^2\left(\sqrt{\frac{1}{2m^2-1}}\xi,m\right),
\end{eqnarray}
\begin{eqnarray}\label{e22}
	u_{iz}=\frac{12\left(B\lambda^2+C\right)}{2-m^2}\mathrm{dn}^2\left(\sqrt{\frac{1}{2m^2-1}}\xi,m\right).
\end{eqnarray}
%Jacobi (sn) solution:
%\begin{eqnarray}\label{e23}
%	u_{iz}=3(u_0-A-D\lambda)sec{h^2}\left(\sqrt{\frac{u_0-A-D\lambda}{4(B\lambda^2+C)}}\xi\right),
%\end{eqnarray}
It is clear that the velocity of the solitary profile depends on  the coefficients $A $ and $D$ and so they  can play important roles in the phase shift of the solitary waves. Furthermore, the amplitude of the profiles of both the solitary and periodic waves can be significantly altered due to the change in the coefficients $B$ and $C$ of the ZK-like equation.  
\section{Results and discussion}\label{5}
\begin{figure} 
	\centering
	\includegraphics[scale=0.45]{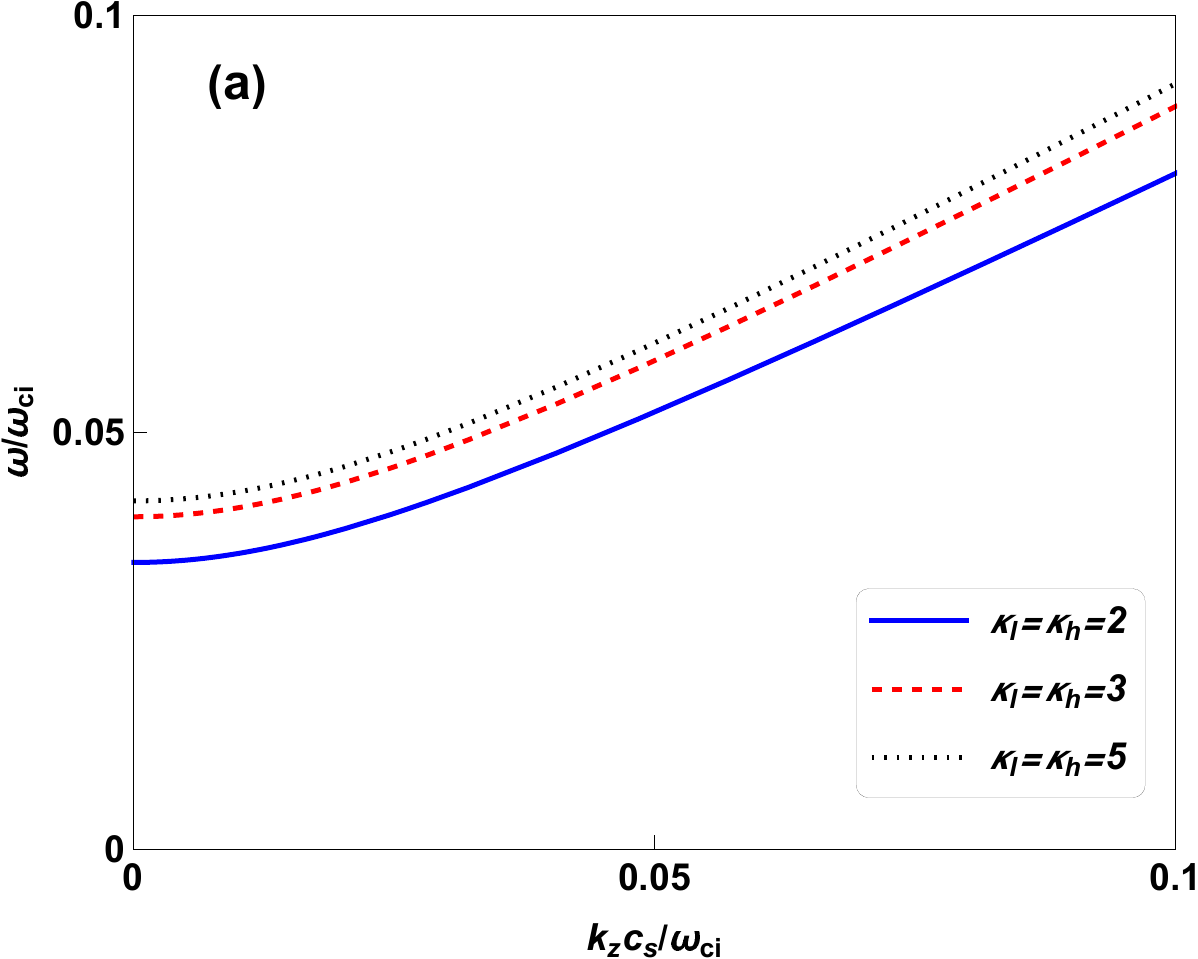}
	\includegraphics[scale=0.45]{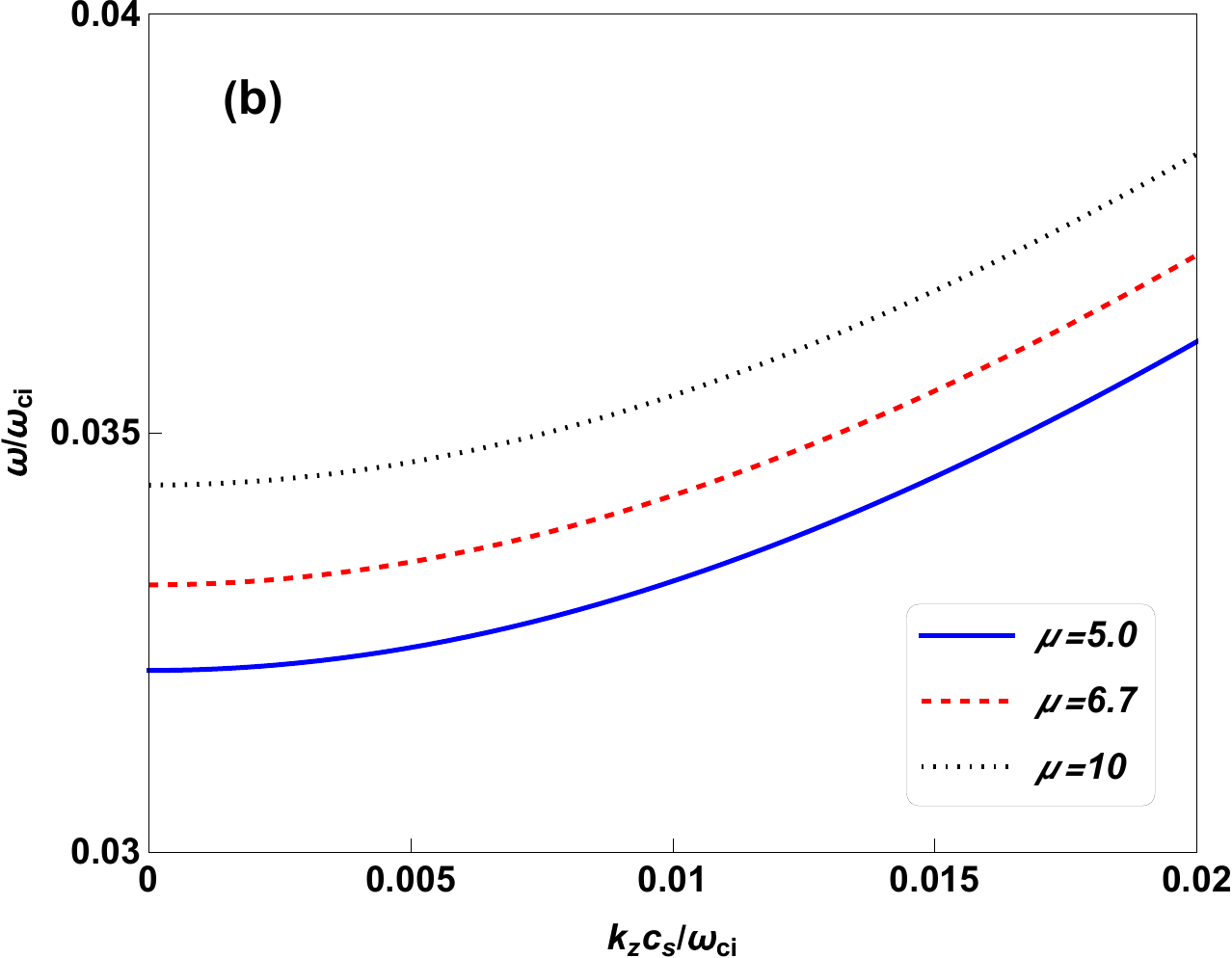}
	\includegraphics[scale=0.45]{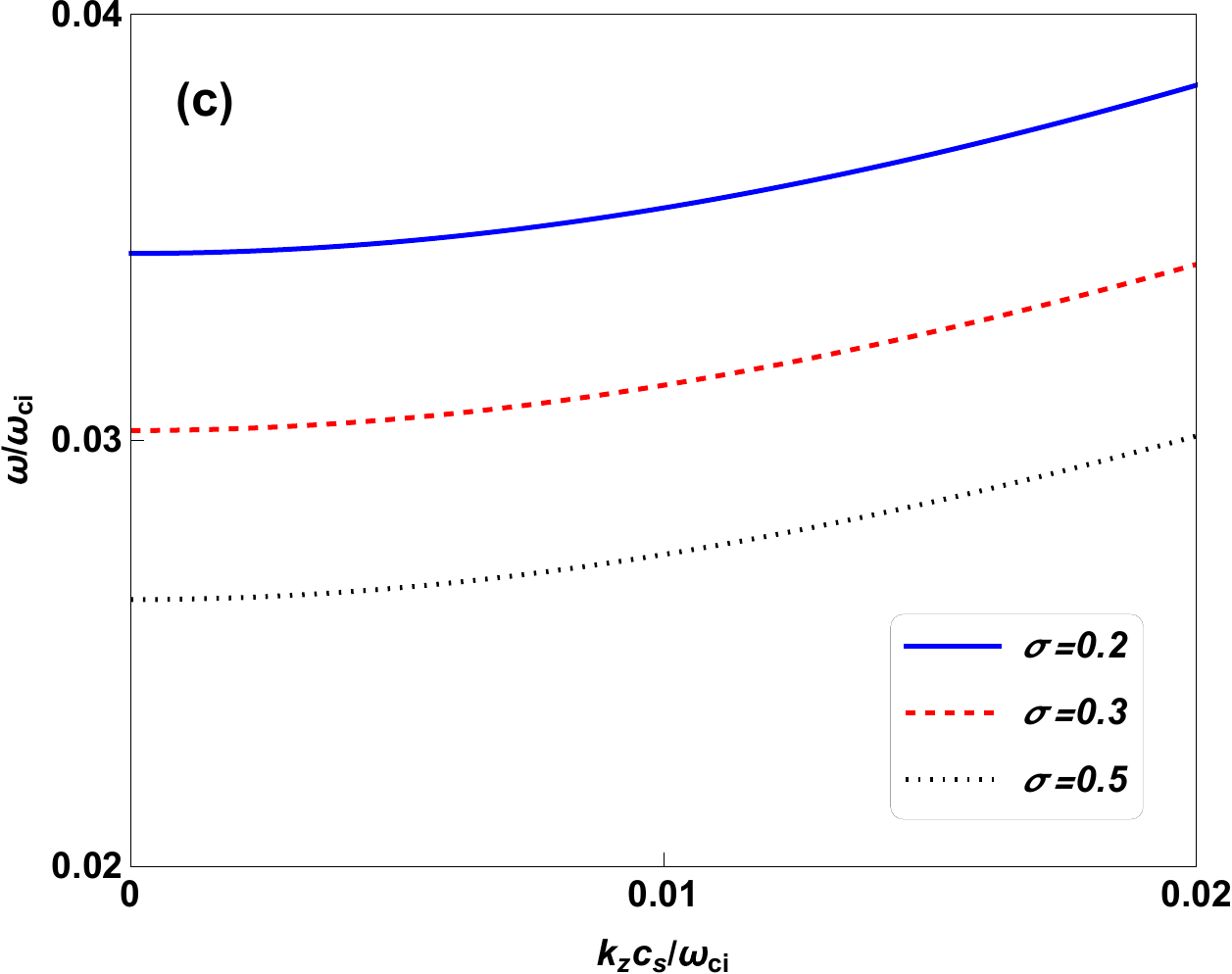}
	\includegraphics[scale=0.45]{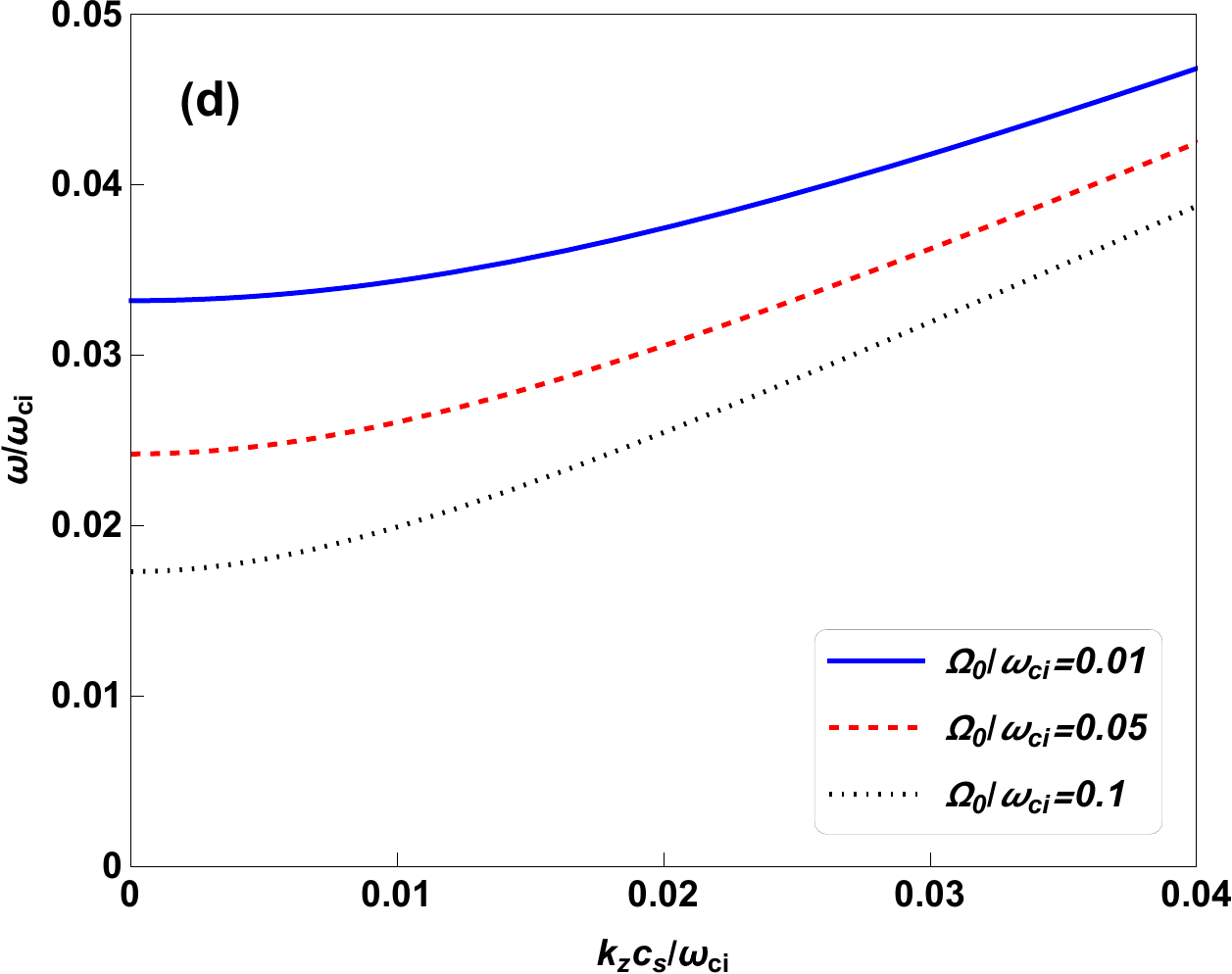}
	\includegraphics[scale=0.45]{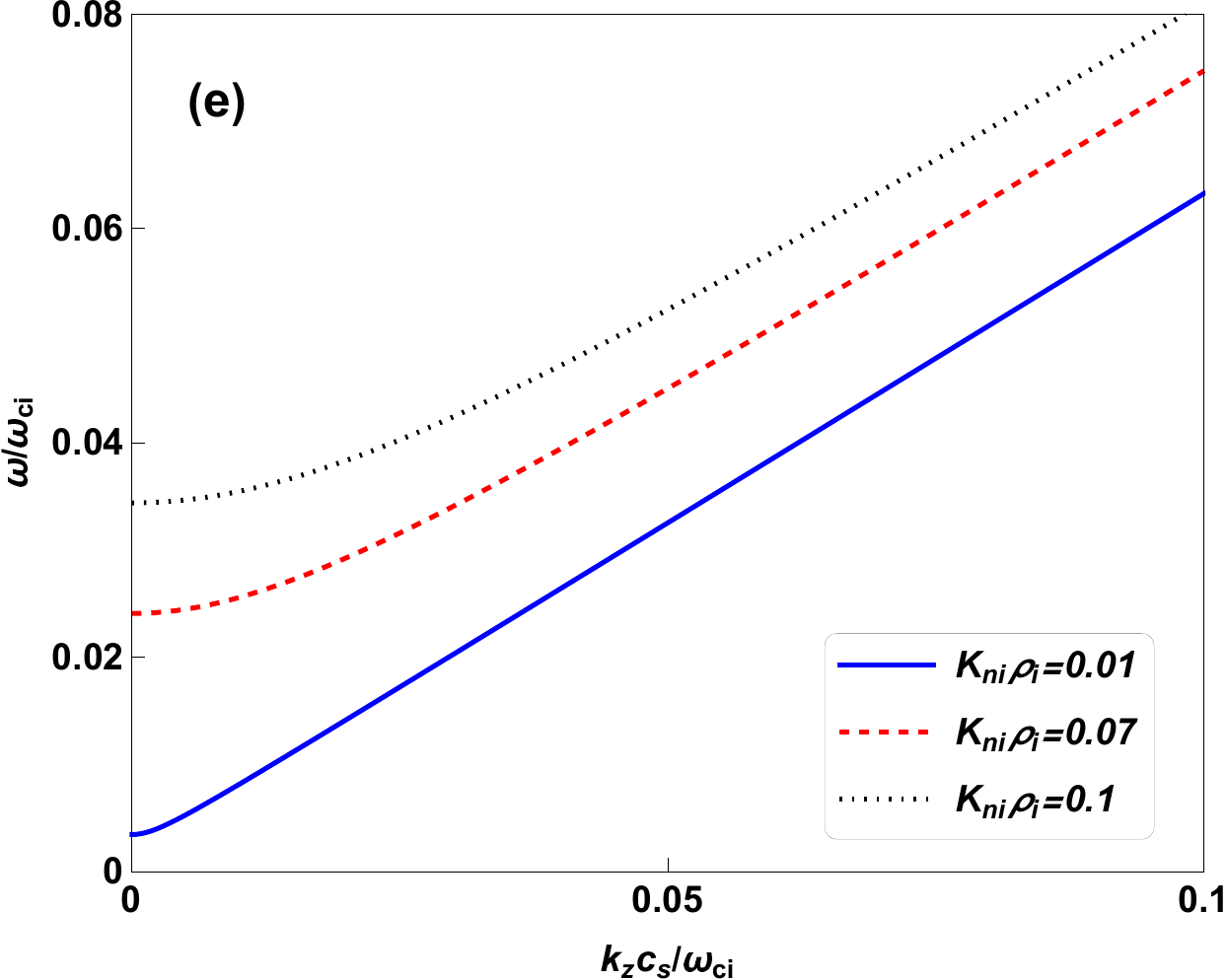}
	\caption{Dispersion curves corresponding to the solutions of Eq. \eqref{e10} are plotted to show the variations of the normalized    frequency  $\omega/\omega_{ci}$  against the normalized wave number  $k_zc_s/\omega_{ci}$ for different values of the  plasma parameters as in the legends.   The other parameter values corresponding to the subplots (a) to (e), respectively, are (a) $k_yc_s/\omega_{ci}=0.1$, $K_{ni}\rho_i=0.07$, $\sigma=0.2$, $\mu=10$, and $\Omega_0/\omega_{ci}=0.05$; (b) $k_yc_s/\omega_{ci}=0.1$, $K_{ni}\rho_i=0.07$, $\sigma=0.2$, $\kappa_l=\kappa_h=2$, and $\Omega_0/\omega_{ci}=0.05$; (c) $k_yc_s/\omega_{ci}=0.1$, $K_{ni}\rho_i=0.07$, $\kappa_l=\kappa_h=2$, $\mu=10$, and $\Omega_0/\omega_{ci}=0.05$; (d) $k_yc_s/\omega_{ci}=0.1$, $K_{ni}\rho_i=0.07$, $\kappa_l=\kappa_h=2$, $\mu=10$, and $\sigma=0.2$; and (e) $k_yc_s/\omega_{ci}=0.1$, $ \sigma=0.2$, $\kappa_l=\kappa_h=2$, $\mu=10$, and $\Omega_0/\omega_{ci}=0.05$.   }     
	\label{fig-linear}
\end{figure}

\begin{figure}
	\centering
	\includegraphics[scale=0.45]{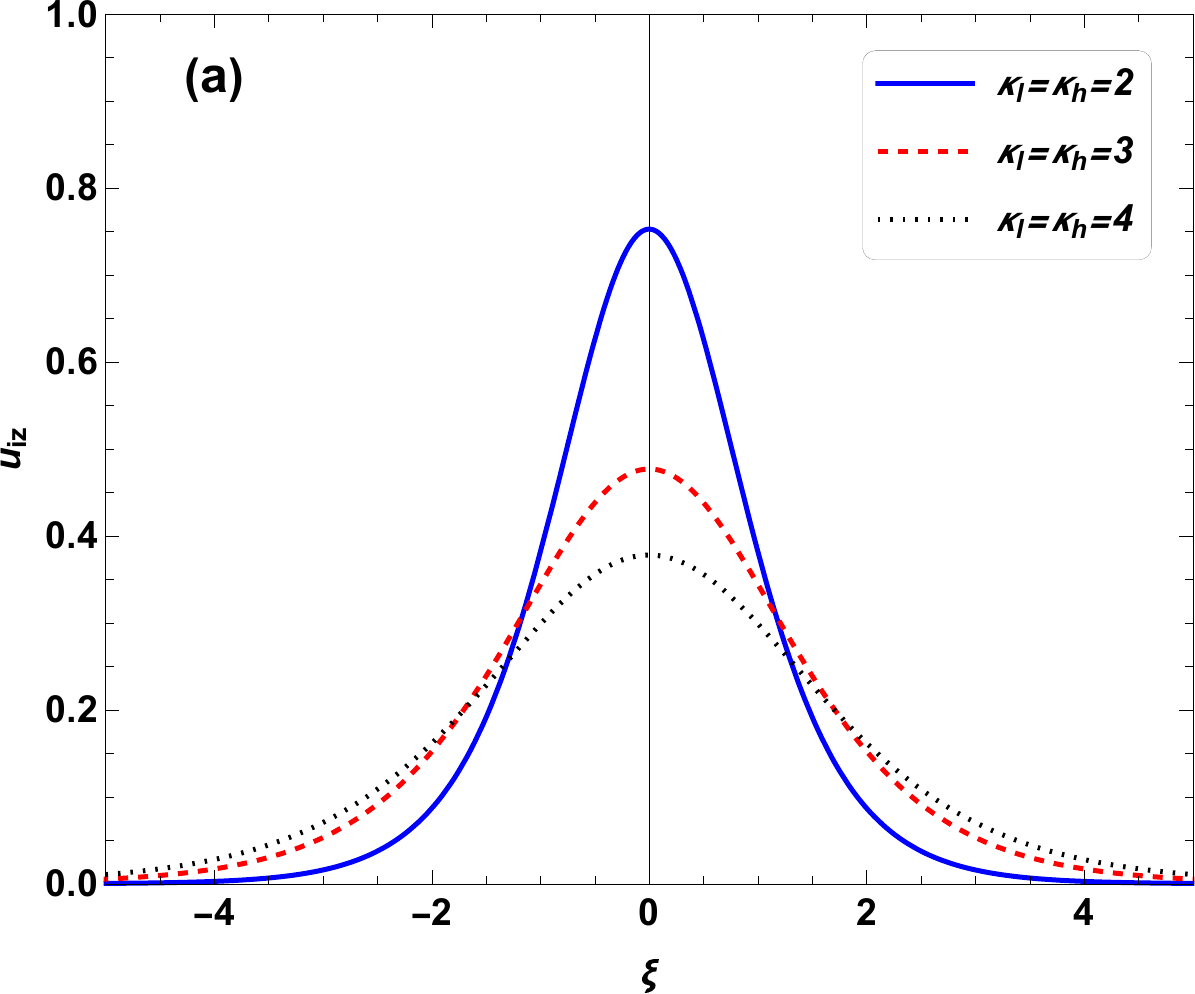}
	\includegraphics[scale=0.45]{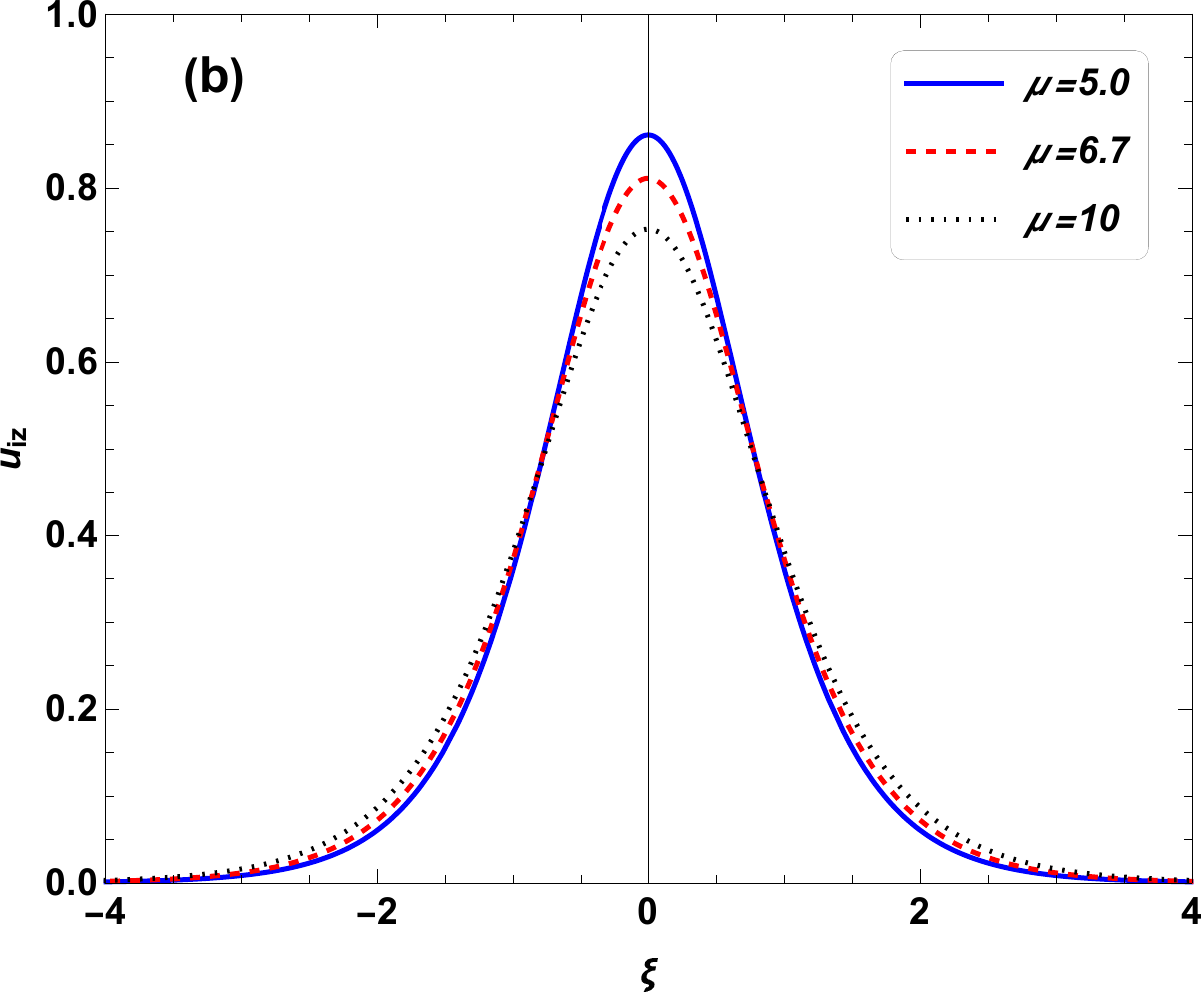}
	\includegraphics[scale=0.45]{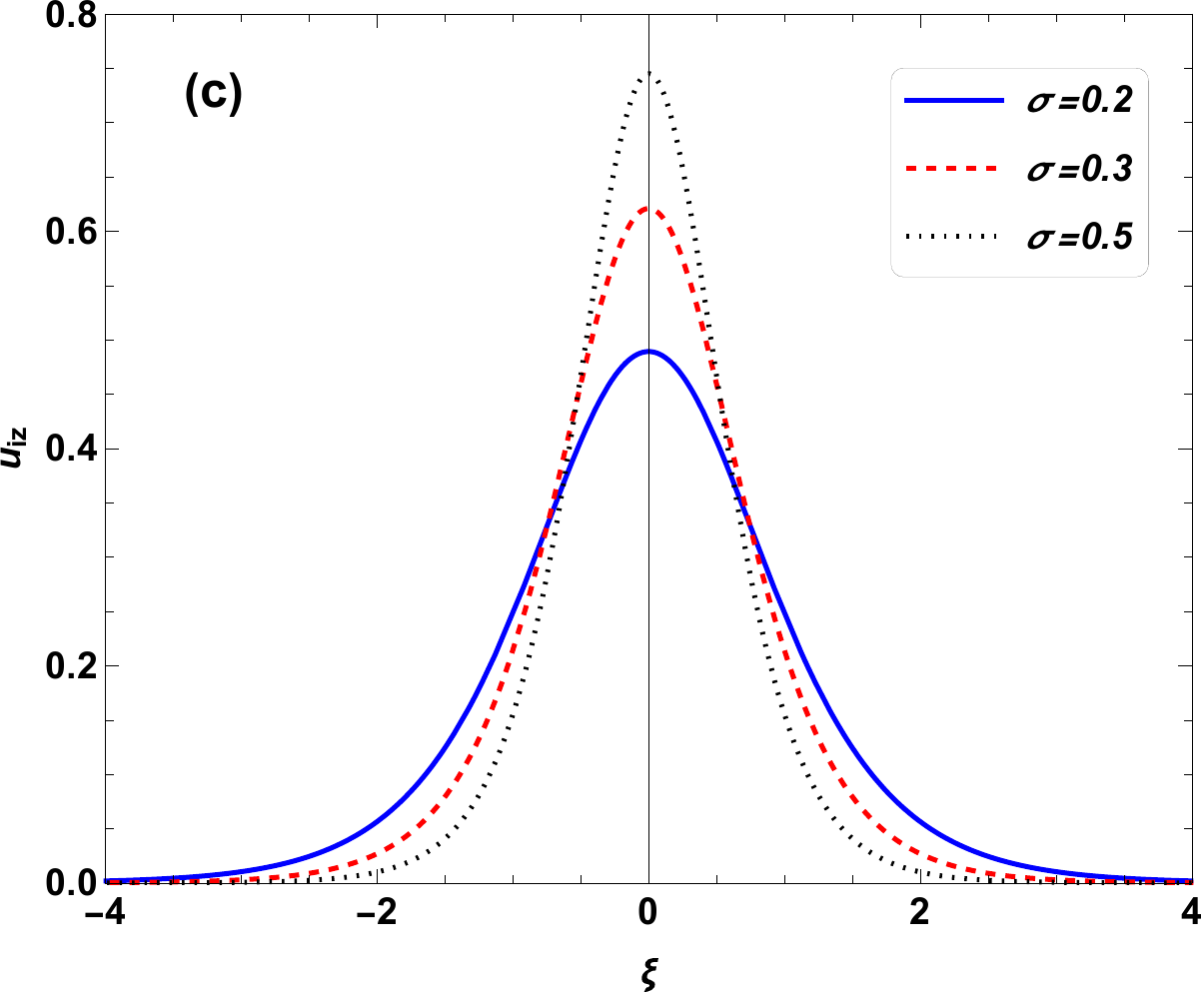}
	\includegraphics[scale=0.45]{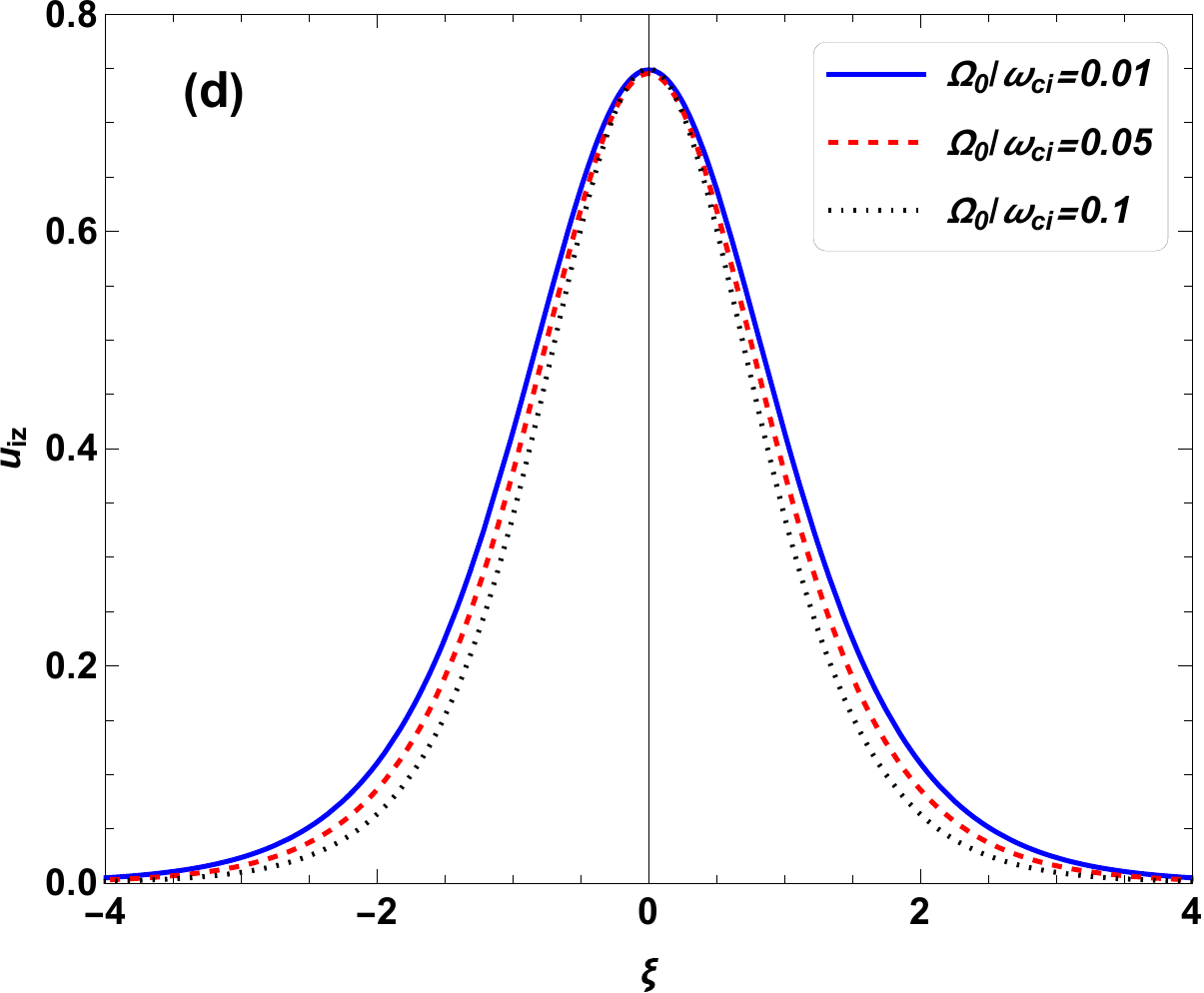}
	\includegraphics[scale=0.45]{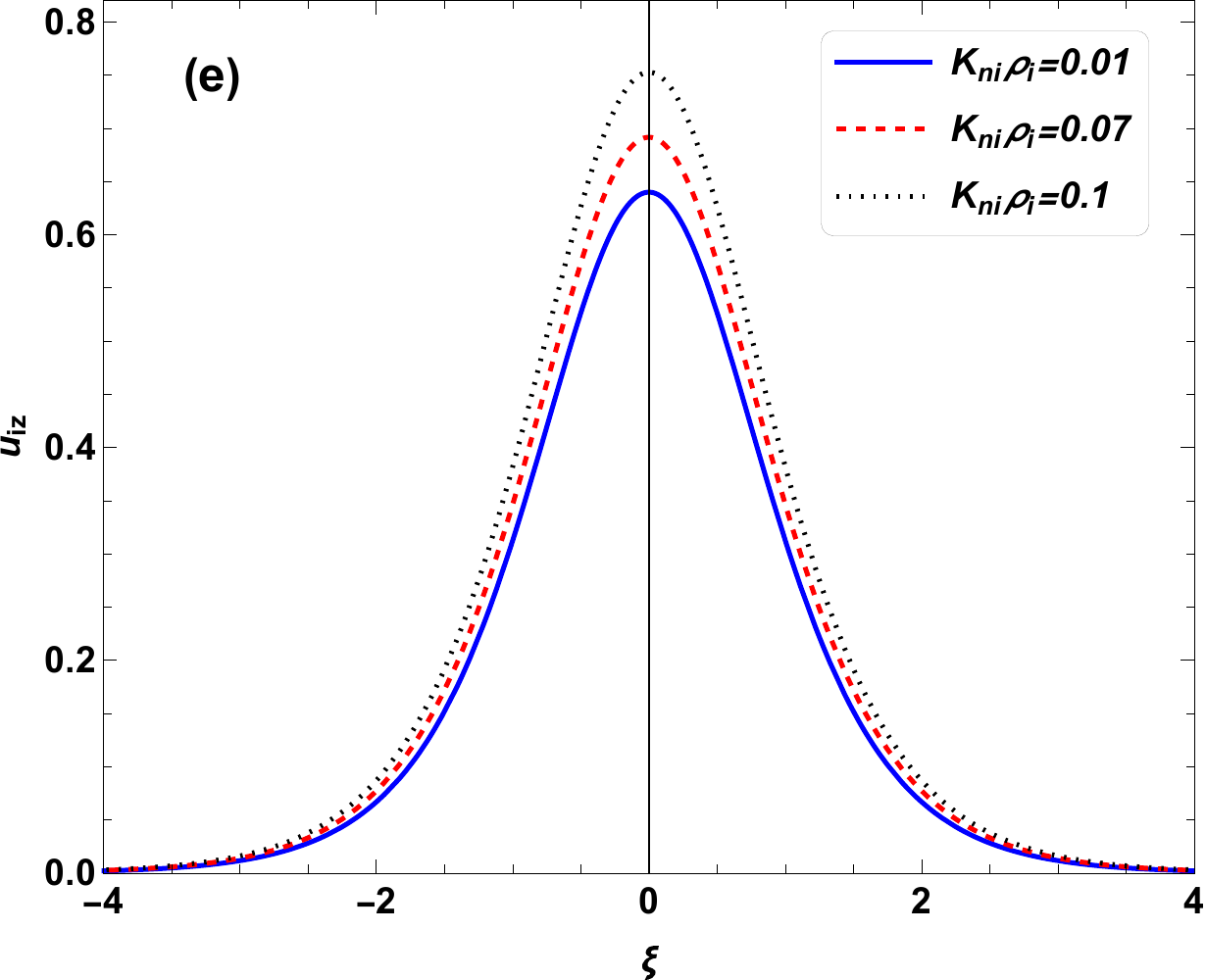}
	\caption{Profiles of the soliton [Eq. \eqref{e20}] are shown for different values of the plasma parameters as in the legends.   The other parameter values for the subplots (a) to (e) are the same as in Fig. \ref{fig-linear}.} 
	\label{fig-solit}
\end{figure}

\begin{figure}
	\centering
	\includegraphics[scale=0.45]{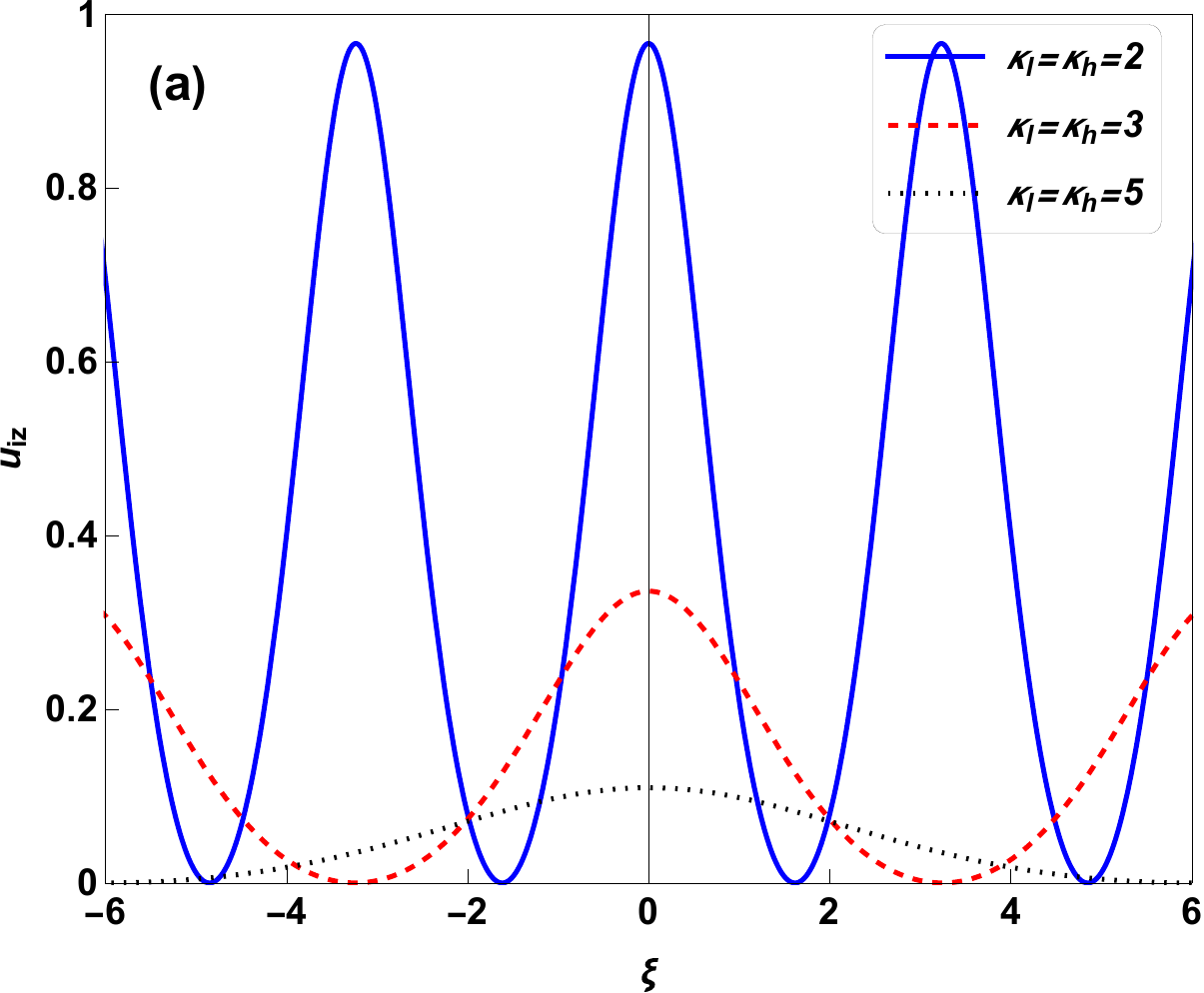}
	\includegraphics[scale=0.45]{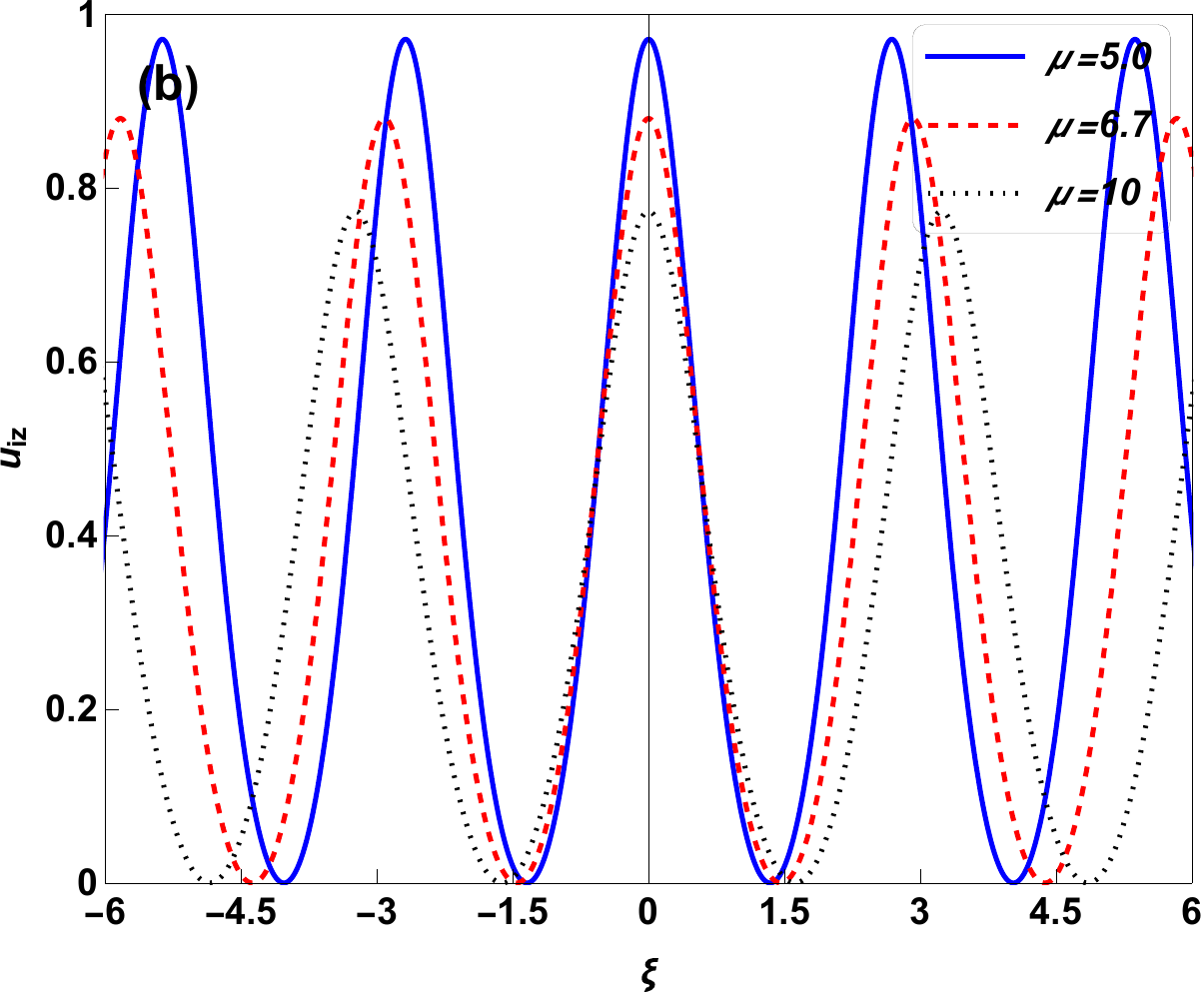}
	\includegraphics[scale=0.45]{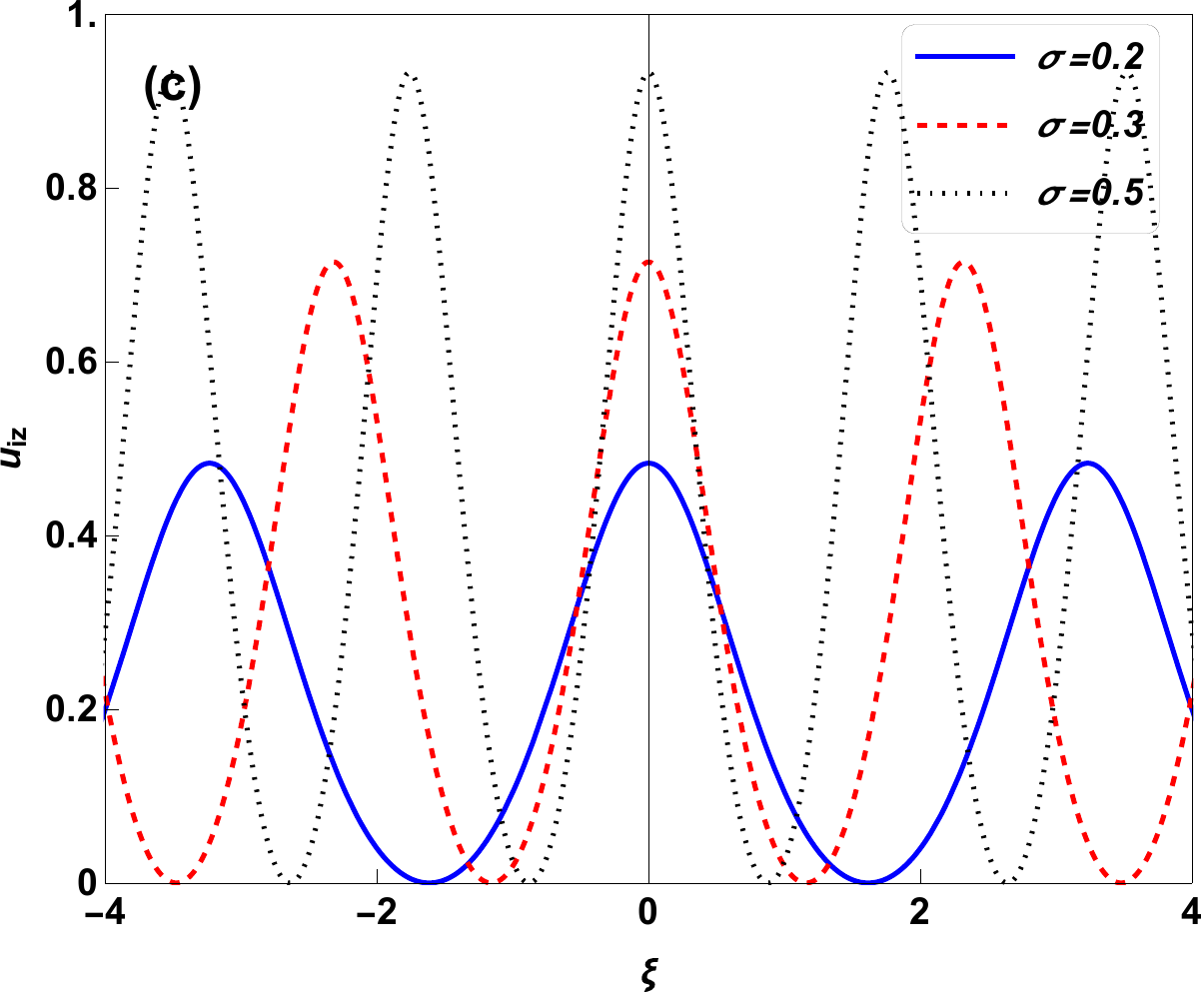}
	\includegraphics[scale=0.45]{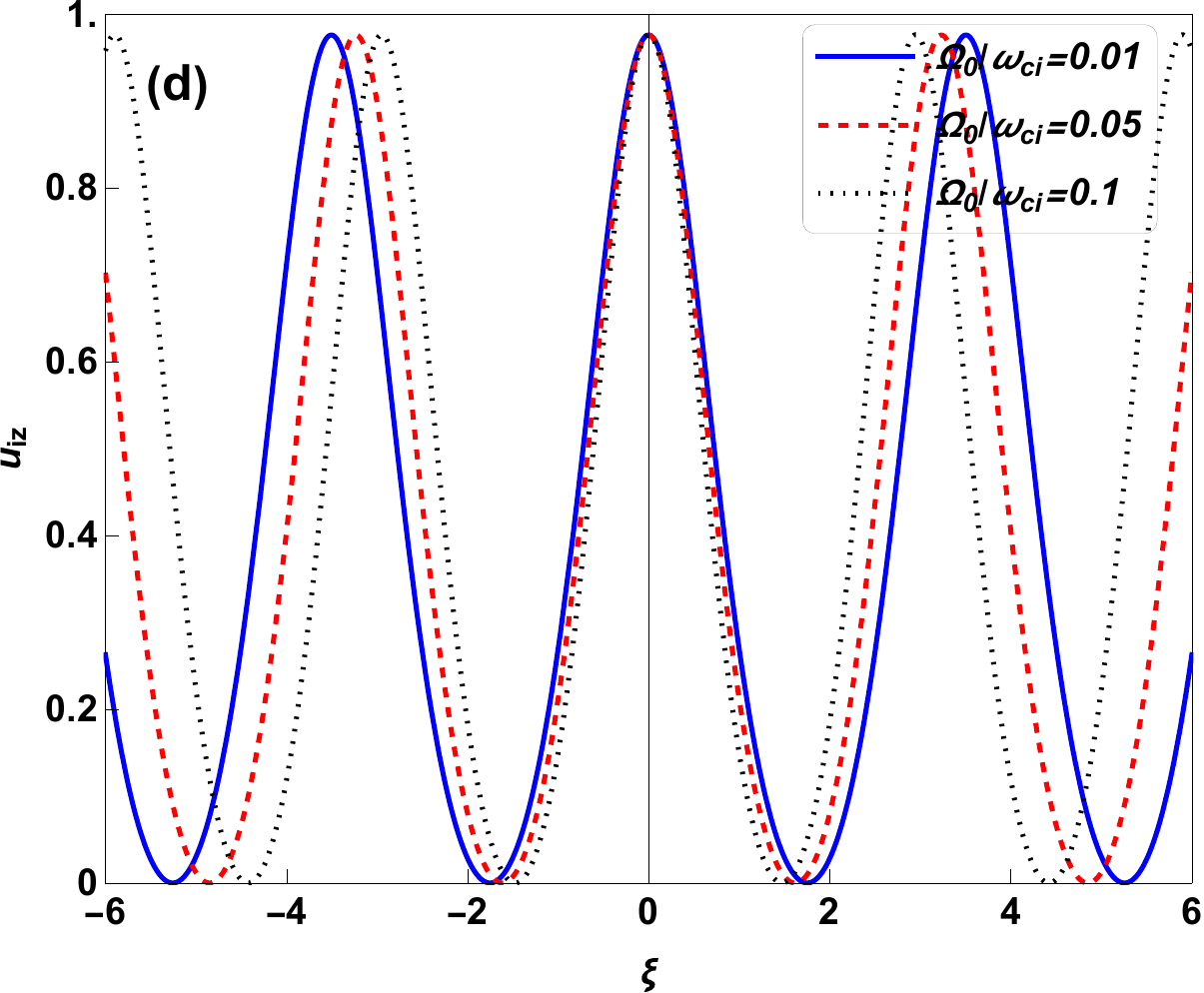}
	\includegraphics[scale=0.45]{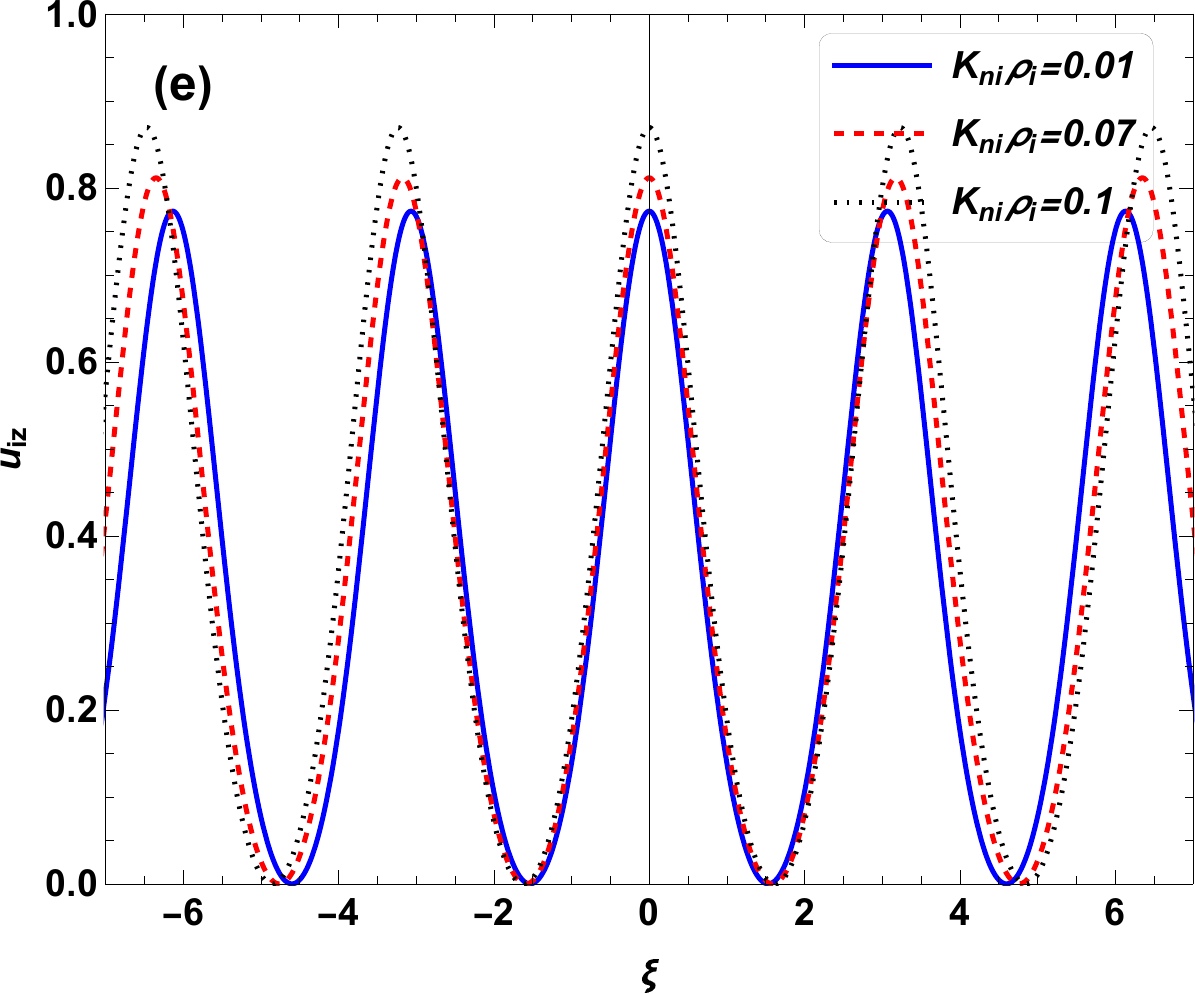}
	\caption{Profiles of the cnoidal wave [Eq. \eqref{e21}] are shown for different values of the plasma parameters as in the legends. The other parameter values for the subplots (a) to (e) are the same as in Fig. \ref{fig-linear}.}
	\label{fig-cn}
\end{figure}

We numerically analyze the effects of some parameters, namely the spectral indexes ($\kappa_l$, $\kappa_h$) associated with the superthermal electrons, the temperature ratio  ($\sigma=T_l/T_h$), the  electron number density ratio ($\mu=n_{h0}/n_{l0})$,   the rotational plasma frequency ($\Omega_0$),  and the density inhomogeneity $(K_{ni})$ on the propagation characteristics of linear drift IAWs as well as on the profiles of the nonlinear solitary and periodic wave solutions of the ZK-like equation \eqref{e18}. To this end, we  choose the parameters that are relevant for space plasmas including those in interstellar plasma medium \cite{BalukuA04227, Schippers, Kaur331, Bukhari11, Singh5504, Sultana581} as $n_{h0}\sim 0.2-10.5~  \mathrm{cm}^{-3}$, $n_{l0}\sim0.02-0.18~ \mathrm{cm}^{-3}$, $\kappa_l\sim 2.1-8$, $\kappa_h\sim 3-8$, $T_l\sim 1.8-10.2~ \mathrm{ eV}$, $T_h\sim 300-1000~ \mathrm{eV}$ and $B_0=2 ~\mathrm{G}$.  In these regimes, the ion gyrofrequency scales as $\omega_{ci}\sim 10^4$ s $^{-1}$. The values of the rotational frequency $\Omega_0$ are typically lower than $\omega_{ci}$. Different values of $\Omega_0$ are considered as in Figs. \ref{fig-linear} to \ref{fig-cn}.  
\par 
Figure \ref{fig-linear} displays the linear dispersion curves for the normalized wave frequency $\omega/\omega_{ci}$ against the normalized wave number $k_zc_s/\omega_{ci}$ for different values of the spectral indexes $\kappa_l$ and $\kappa_h$; the density ratio $\mu$, the temperature ratio $\sigma$,   the rotational frequency $\Omega_0$,  and the density inhomogeneity $K_{ni}$. It is found that the frequency of the wave increases with increasing values of   the wave number $k_z c_s/\omega_{ci}$  as well as the  parameters $\kappa_l$ or $\kappa_h$, $\mu$ and $K_{ni}$  except that the frequency decreases with increasing values of  $\sigma$  and $\Omega_0/\omega_{ci}$. Such an increase or decrease of the frequency becomes significant even with a small   change of the parameter values.  From the subplot \ref{fig-linear}(a), it is also noticed that a further enhancement of the spectral indexes beyond $\kappa_l=\kappa_h=5$ (in which case one approaches the Boltzmann distribution of electrons) will not result into any significant increase of the wave frequency.    Since the dispersion relation [Eq. \eqref{e10}] generalizes some previous ones, one can calculate the frequency in different cases including that of the pure drift waves at $k_z = 0$. Physically, increasing  the values of the  plasma parameters ($\kappa_l$ or $\kappa_h$, $\mu$ and $K_{ni}$)  results into an increase of the restoring force,  which shortens the oscillation period and increases the frequency \cite{Ahmad905810505}.  It also follows that  when the percentage of energetic electrons becomes higher or the length scale for the density inhomogeneity is lower, the drift IAWs having lower phase velocity may be weakly damped or undamped in the wave-particle interaction processes.   
%%%%%%%%%%%%%%
%%%%%%%%%%%%%%%%
%%%%%%%%%%%%%%%%
\par 
Figure \ref{fig-solit} shows the profiles of the solitary wave solution \eqref{e20} for different values of the plasma parameters as mentioned before.  It is found  that while the amplitude decreases, the width of the solitary wave increases with increasing values of either $\kappa_l$ or $\kappa_h$ and $\mu$ [Subplots \ref{fig-solit}(a) and \ref{fig-solit}(b)]. This is due to the fact that  the amplitude of the solitary wave is inversely proportional to the spectral indexes  $\kappa_l$ and $\kappa_h$, and the  density ratio $\mu$. Physically, the higher superthermality or hotter species of electrons stipulates the existence of electrons with higher thermal energies. In this situation, more ions need to be clustered in order to construct the Debye sheath which is almost impossible since the ions are much less mobile than electrons. So, in order to sustain the solitonic structure, its amplitude must be inversely proportional to the spectral index parameter and the high-to-low electron density fraction. From the subplots \ref{fig-solit}(c) and \ref{fig-solit}(d), it is also seen that in contrast to the effects of the increasing values of $\sigma$ by which the soliton amplitude increases, however,  the width decreases,   the amplitude remains the same for increasing values of the rotational frequency $\Omega_0$ and the width increases. These occur since similar to the effects of  $\mu$, the parameter  $\sigma$ is also inversely proportional to the amplitude of the solitary waves. 
 It is noteworthy to mention that in plasmas, the charged particles are dispersed due to the convection. The tendency to move  towards the rarefied regions from the compressed regions  is persuaded due to the thermal motions of ions where  the slowness of  them disables the Debye sheath not to complete.  Since the soliton  width is inversely proportional to $\Omega_0$, it follows that    the larger the values of the rotational  frequency, the reductions are the dispersion and the wave energy of solitons.  Furthermore, from the subplot \ref{fig-solit}(d) it is seen that both the amplitude and width of the soliton can be increased with an increasing value  of $K_{ni}~(\sim\rho^{-1}_i)$. Physically,  an enhancement of $K_{ni}$ is related to  a   reduction of the   length scale of the density inhomogeneity which, in turn, increases  the wave dispersion and hence both the amplitude and width of the soliton.       
%%%%%%%%%%%%%%%%%%%%%%%%%%%%%%%%%%%%%%%
%%%%%%%%%%%%%%%%%%%%%%%%%%%%%%%%%%%%
\par 
As an illustration, we plot the profiles of the cnoidal wave  [Eq. \eqref{e21}] as shown in Fig. \ref{fig-cn}. It is seen that  the qualitative features with the variations of the plasma parameters remain unaltered as those for solitary solutions [Fig. \ref{fig-solit}]. 
%%%%%%%%%%%%%%%%%%%%%%%%%%%%%%%%%%%  
%%%%%%%%%%%%%%%%%%%%%%%%%%%%%%%%%%%%%
\section{ Conclusions}\label{6}
We have investigated the dispersion properties and the nonlinear evolution of coupled drift waves and ion-acoustic waves in a nonuniform rotating magnetoplasma with two-temperature superthermal electrons. It is found that the density gradient couples the drift mode and the ion-acoustic wave. In the linear limit, a frequency enhancement of the drift IAW is found as one approaches from superthermal distribution to Maxwellian distribution of electrons with increasing values of the spectral indexes $\kappa_l$ and $\kappa_h$.   The frequency is also enhanced with an increase of the density ratio, however, with a  reduction of any of the length scale of the density inhomogenity, the temperature ratio and the rotational frequency.     Such an enhancement of the wave frequency and hence that of the wave phase velocity may lead to a strong damping of drift IAWs in the wave-particle interactions. Thus, it is expected that deviating from the thermal equilibrium, the plasmas with superthermal electrons can support weak wave damping \cite{chatterjee2015}. On the other hand, in the intermediate regions between the two domains of the wave number at which the low- and high-frequency branches of  drift IAWs can be recovered, a ZK-like evolution equation is derived using the dispersion approach. By the Jacobi function expansion technique, new traveling wave solutions, namely solitary and periodic (cnoidal and dnoidal) solutions are  obtained and their properties are studied by the effects of the plasma parameters, namely the electron superthermality spectral indexes ($\kappa_l$ and $\kappa_h$); the high to low electron density fraction ($\mu$), the ratio of the low to high electron temperatures ($\sigma$),   the rotational plasma frequency ($\Omega_0$),  and the plasma density gradient $K_{ni}$. It is found that the  profiles of the solutions are significantly altered by the effects of these parameters. 
\par 
In conclusion, the dispersion properties of drift ion-acoustic waves and the possibilities of the formations of nonlinear solitary and periodic wave structures should be useful for understanding the salient features of drift ion-acoustic perturbations   that may spontaneously emerge in a nonuniform, rotating magnetized    space plasmas that contain two-temperature superthermal electrons,  and thereby giving rise to the dominant mechanism for transport of particles, energy and momentum across the magnetic field lines. 
 
\section*{Acknowledgments}
 The authors thank the anonymous referees for their useful comments.  A. P. Misra  acknowledges   support from the Science and Engineering Research Board (SERB, Government of India) for a research project with sanction order no. CRG/2018/004475.

\renewcommand{\theequation}{A-\arabic{equation}}
\setcounter{equation}{0}  % reset counter 

\appendix
\section{Derivation of the Coriolis and centrifugal forces \label{appendixA}}

In Newtonian mechanics,  the scalar coefficients of a vector $\mathbf{A}=A_x\hat{i}+A_y\hat{j}+A_z\hat{k}$ remain invariant for any given frame of reference, i.e.,
\begin{equation}
\frac{dA_{x,y,z}}{dt}\Big|_I=\frac{dA_{x,y,z}}{dt}\Big|_R,
\end{equation} 
where $I$ and $R$ stand for the values of the scalar quantities in the inertial  and  rotating frames respectively. Next, using the product rule of differentiation and  the definition of the angular velocity of a particle at a position, which gives $d\hat{i}/dt=\mathbf{\Omega}\times\hat{i}$ etc., we obtain
\begin{equation}
\frac{d\mathbf{A} }{dt}\Big|_I=\frac{d\mathbf{A}}{dt}\Big|_R+\mathbf{\Omega}\times \mathbf{A}.
\end{equation} 
So, if $\mathbf{r}$ is the position vector of a point on the surface of a rotating body, we have the velocity,
\begin{equation}
\mathbf{v}_I\equiv\frac{d\mathbf{r}}{dt}\Big|_I=\frac{d\mathbf{r}}{dt}\Big|_R+\mathbf{\Omega}\times \mathbf{r},
\end{equation}
and similarly the acceleration,
\begin{equation}
\frac{d\mathbf{v}_I }{dt}\equiv \frac{d^2\mathbf{r} }{dt^2}\Big|_I=\frac{d^2\mathbf{r} }{dt^2}\Big|_R+2\mathbf{\Omega}\times\frac{d\mathbf{r} }{dt}\Big|_R+\mathbf{\Omega}\times(\mathbf{\Omega}\times\mathbf{r}).
\end{equation}
Thus, using the Newton's second law of motion, the net force acting at a position $\mathbf{r}$ on the surface of rotating body in a rotating frame of reference is give by
\begin{equation}
\mathbf{F}_R=m\frac{d^2\mathbf{r} }{dt^2}\Big|_R=\mathbf{F}_I-2m\mathbf{\Omega}\times \mathbf{v}_I-m\Omega^2\mathbf{r},
\end{equation}
where $\mathbf{F}_I=m\frac{d^2\mathbf{r} }{dt^2}\Big|_I$ is the force in the inertial frame; the second and the third terms on the right-hand side are, respectively, the Coriolis force and the centrifugal force. \\

\bigskip
\noindent

\end{document}